\documentstyle[prd,aps,preprint,tighten,floats]{revtex}

\newcommand{\rf}[1]{(\ref{eq:#1})}

\date{November 5, 1996}
\begin{document}

\author{ M. D. Maia\thanks{E-Mail: maia@mat.unb.br}}
\address{Universidade de Bras\'{\i}lia, Departamento de Matem\'{a}tica\\ 
Bras\'{\i}lia, DF. 70910-900}
\title{About Time} 
\maketitle
\draft
\preprint{UnB.FM.M.-003.96}
pacs{04.20.Fg, 04.60.+n, 12.25.+e}
\begin{abstract}
The concept of time  is discussed in the context of the canonical formulation
of the gravitational field.  
Using  a hypersurface orthogonal foliation, the arbitrariness
of the lapse function is eliminated  and the  shift vector
vanishes, allowing  a consistent  definition of time. 
\end{abstract}

\section{On Lapses and Shifts}
As it well know, the space-time of Newtonian
mechanics is foliated  by  globally defined
3-dimensional simultaneous sections which  characterizes a
hypersurface  orthogonal  propagation vector field. This is  mainly  a
consequence of the Galilei symmetry and the  absolute time
may be  described  as  foliation time,  defined by the affine parameter of an
integral curve of that  vector  field. 
Likewise, in  special relativity a local foliation of Minkowski  space-time
may be introduced as a sequence of 
light cones,  whose structure is invariant under the Poincar\`e group.
Then the  local time of special relativity may  be  also defined as  a  
parameter (proportional to the arclength) of the  integral curves  of  the
propagation vector  of that  foliation. 

When we move to general relativity,  the  Poincar\`e group is replaced by the
manifold mapping group and  time is usually taken to be just
one of the local  coordinates.  When this is combined with the principle 
of general covariance, it becomes very difficult to
characterize a  clear notion of time, specially  when  working 
with problems which  demand  an explicit   time
parameter, such as the canonical formulation of  general
relativity\cite{Unruh:1}.    
 The purpose of this note is to  reexamine concept of time in that context 
 and  to propose an appropriate foliation  which may be used
to  define a  coordinate independent notion of time.
We  start with a brief  review of the  standard Dirac-ADM  
canonical formulation  of the gravitational field \cite{Dirac:1},\cite{ADM}.

Given a space-time ${\cal V}_{4}$, with metric components ${\cal
G}_{\alpha\beta}$ in arbitrary coordinates, at
each point it may be locally decomposed into a space-like
3-dimensional hypersurface $\Sigma(t)$  with metric $h_{ij}$,   and a
time-like vector field\footnote{We use 
the following notation:  Small case Latin indices refer to a
3-surface  and run from 1 to 3. Greek indices refer to  the
four dimensional  space-time  ${\cal V}_{4}$, running from 0 to 3.
The 3 dimensional metric is  denoted by $h_{ij}$  and 
$\nabla_{i}$  denotes  the corresponding covariant derivative.
The covariant derivative with respect to the  4 dimensional metric is
denoted by the usual semicolon. The signature  of the space-time is
$-+++$  and  its  Riemann curvature  is
${\cal R}_{\alpha\beta\gamma\delta}$, with Ricci curvature
${\cal R}_{\beta\gamma}={\cal G}^{\alpha\delta}{\cal
R}_{\alpha\beta\gamma\delta}$.  The  3 dimensional curvature tensors  are
denoted by  capital $R's$: $R_{jk} = h^{il}R_{ijkl}$ and $R=h^{ij}R_{ij}$ }
$\zeta^{\mu}$ :    
\[
\zeta^{\alpha}\zeta_{\alpha}={\cal G}^{00}\zeta_{0}^{2}+2
{\cal G}^{0i}\zeta_{0}\zeta_{i}+{\cal G}^{ij}\zeta_{i}\zeta_{j}=-1 .
\]
Using  coordinates such that  ${\cal G}^{00}\zeta_{0}^{2}=-1,\;\,$  
$2{\cal G}^{0i}\zeta_{0}\zeta_{i}+{\cal G}^{ij}\zeta_{i}\zeta_{j}=0$ and
defining  the {\em lapse} function $N=\zeta_{0}$ and  the
3-dimensional {\em shift} vector  $N_{i}=\zeta_{i}$,  the metric of ${\cal
V}_{4}$ may be written as

\begin{equation}
{\cal G}_{\alpha\beta}= \left(
\begin{array}{cc}
-(N^{2}-N^{m}N_{m})   & N_{i}\\                
               N_{j}      &  h_{ij}   
  \end{array}   \right)
\;\;\;\; \mbox{and} \;\;\;\;  
{\cal G}^{\alpha\beta}= \left(   
\begin{array}{cc}
-1/N^{2}      &  N^{i}/N^{2}\\  
N^{j}/N^{2} & -(h^{ij}-N^{i}N^{j}/N^{2}) 
\end{array}  \right) 	\label{eq:ADM}
\end{equation}
The extrinsic curvature of  $\Sigma(t)$,  $K_{ij} $ is defined by the covariant
derivative of the unit normal  vector to $\Sigma(t)$, using  affine connection
of  $h_{ij}$.  

By direct calculation of the Ricci scalar ${\cal R}$ of ${\cal V}_{4}$, in the
above  parametrization the  Einstein-Hilbert Lagrangian in  the ADM  form
may be written as
\begin{equation}
{\cal L}_{ADM} = {\cal R}\sqrt{{\cal
G}}=-(K^{ij}-Mh^{ij})\dot{h}_{ij}\sqrt{h}-N H^{0} 
-N^{i} H^{i}- 2\frac{d}{dt}(M\sqrt{h})-\nabla_{i}\varphi^{i} ,\label{eq:LADM}  
\end{equation}
where    $M=h^{ij}K_{ij}$,  $K^{2}=K^{ij}K_{ij}$ and where
\begin{eqnarray}
H^{0}=-\left[ R+ (M^{2}- K^{2})\right] \sqrt{h}, \label{eq:constraint1}
\vspace{3mm}\\
H^{i} =2\nabla_{j}\left[ K^{ij}-Mh^{ij}\right]. \sqrt{h} \label{eq:constraint2}
\end{eqnarray}
The last term in \rf{LADM} is  a total divergence of the vector
\[
\varphi^{i} = 2\left[ (K^{ij}+Mh^{ij})N_{j}-(M N^{i}-
\nabla^{i}N) \sqrt{h})\right].
\]
so that it may be discarded  in the variational process, provided  $\Sigma(t)$
is  compact. 
On the other hand, the total time derivative term in \rf{LADM} does not
contribute to the field 
equations and it may be also removed by canonical transformations.
Therefore discarding  these two terms,  the effective
Lagrangian may be  written simply as  
\[
{\cal L}_{ADM} =\pi^{ij}\dot{h}_{ij} -{\cal H}_{ADM},
\] 
where  we have denoted   ${\cal H}_{ADM}=N H^{0}+ N_{i}H^{i} $ and 
$\pi^{ij}$ is the  momentum canonically conjugated to $h_{ij}$, given by
\begin{equation}  
\pi^{ij}=\frac{\partial {\cal L}_{ADM}}{\partial \dot{h}_{ij}}=-
(K^{ij}-Mh^{ij})\sqrt{h}. \label{eq:pi}
\end{equation}
Therefore, ${\cal H}_{ADM}$ is the  Hamiltonian of the system.
Noting that  $\pi^{ij}\pi_{ij}=(K^{2}+M^{2})h $ and denoting  $ \pi
= h_{ij}\pi^{ij} =2M\sqrt{h}$ this  Hamiltonian may also be expressed as
\begin{equation}
{\cal H}_{ADM}= -N[ R +\frac{1}{h}(\frac{\pi^{2}}{2} -\pi^{ij}\pi_{ij})]
\sqrt{h} -2N_{i}\nabla_{j} \pi^{ij}. \label{eq:H}
\end{equation}
The rest of this  story  is too well known: Taking the variation of the action
with respect to $N$ and $N_{i}$  gives the superhamiltonian  constraint
$H^{0}=0$ and the  supermomentum constraint $H^{i}=0$, so that ${\cal H}_{ADM}$ vanishes. 
As long as we remain in the classical side of the  theory, the system may be
solved as a  Dirac's constrained system.
However, in the quantum side of the theory things  get  more
complicated. First of all because
when we translate the Hamiltonian  as  an operator  $\hat{{\cal
H}}_{ADM}$, acting on the physical
Hilbert space, then its eigenfunctions
become frozen in time:  $\frac{\partial \Psi}{\partial t}=
0$ \cite{Komar:1}. Furthermore,  since the  constraints
equations  are used concomitantly with the equations of motion,
the  operator ordering becomes untreatable to say the least, 
 as the Lie algebra of the commutators will not close properly.
This has become known as the  time problem in general
relativity.\cite{Isham,Alvarez,Kuchar}.  
As it has been noted by several authors 
the  ADM  formalism may be written in terms of  foliations. Indeed,
each 3 dimensional hypersurface $\Sigma(t) $ may be characterized by the local
embedding ${\cal X} :   
\Sigma(t) \rightarrow   {\cal V}_{4}$, such  that  for  a given  vector
$\zeta^{\mu}=(N^{0},N^{i})$  we have:
\begin{equation}
h_{ij}={\cal X}^{\mu}_{,i}{\cal X}^{\nu}_{,j}{\cal G}_{\mu\nu},\;\;
{\cal X}^{\mu}_{,i}\zeta^{\nu}{\cal G}_{\mu\nu}=N_{i},\;\;
\zeta^{\mu}\zeta^{\nu}{\cal G}_{\mu\nu}=-(N^{2}-N^{i}N_{i}).\label{eq:embeddX}
\end{equation}
These equations  describe what may be called  a   {\em transverse} foliation of
the space-time.
As such,  it characterizes time as  a  parameter of the diffeomorphism group
defined by the  propagation vector $ \zeta^{\mu}=(N^{0},N^{i})$.
However,  such time  is not  necessarily coincident or compatible with the time
parameter (the coordinate time) already included in \rf{embeddX}. In fact,   the  existence of the  components
$N^{i}$   means that a  coordinate transformation in a given leaf $\Sigma (t)$
may eventually change  the propagation vector
$\zeta^{\mu}$ and consequently the  foliation  defined by \rf{embeddX}.
On the other hand,   the  lapse function $N$ remains an arbitrary function
during the  evolution of the system. Since this  is  essentially a
clock for the coordinate time,  different observers may set this clock at will
and independently.
To make things  even more complicated, the arbitrariness in the sign of
$N^{2}$  
 may induce a classical change in  the space-time signature \cite{Ellis:1}.    
Therefore, while the  coordinate time
remains intact, the  foliation time  may change or even vanish. 
Consequently, although  a transverse foliation such as \rf{embeddX} is
mathematically sound 
(see eg. \cite{Dray:1}),  it is not an  appropriate instrument to  define time in
canonical gravity, at least  in the context of the ADM formulation. 
In the next section  an alternative foliation of the
space-time will be  considered, where the lapse is fixed and the
shift vanishes. 

\section{Shiftless Foliations}
Consider  a  3-surface $\bar{\Sigma}$  taken
as background hypersurface and a time-like unit vector
$\bar{{\cal N}}^{\mu}$ {\em orthogonal} to $\bar{\Sigma}$. 
The  isometric orthogonal embedding of the 3-surface   $\bar{{\cal Y}} :
\bar{\Sigma}\rightarrow  
{\cal V}_{4}$  satisfy the conditions\footnote{All objects  defined
in the background hypersurface  are  denoted with  a  overbar. Since  
 $\bar{\cal Y}^{\mu}$  are scalars with respect the geometry of $\bar{\Sigma}$
their covariant derivatives relative to $\bar{h}_{ij}$ are written simply as
a  colon.}: 
\begin{equation}
\bar{h}_{ij}=\bar{{\cal Y}}^{\mu}_{,i}\bar{{\cal Y}}^{\nu}_{,j}{\cal G}_{\mu\nu},\;\;
\bar{{\cal Y}}^{\mu}_{,i}\bar{{\cal N}}^{\nu}{\cal G}_{\mu\nu}=0,\;\;
\bar{{\cal N}}^{\mu}\bar{{\cal N}}^{\nu}{\cal G}_{\mu\nu}=-1,
\label{eq:embeddY} 
\end{equation} 
The   second fundamental form of $\bar{\Sigma}$  describes the variation of
the  normal vector field  $\bar{\cal N}$  when  its foot displaced  along
$\bar{\Sigma}$, using   the affine connection of the space-time
metric\footnote{Since the metric  connection of  $\Sigma$ is  induced by that
of  ${\cal 
V}_{4}$, this second form is equivalent to the  extrinsic curvature used in the
ADM  formalism and they coincide in the reference frame defined by
$\Sigma$  (se eg. \cite{Isham,Hojman}).}. In terms of the foliation
coordinates it is given by  \cite{Eisenhart:1}
\begin{equation}
\bar{b}_{ij}= -\bar{\cal Y}^{\mu}_{,i}\bar{\cal N}^{\nu}_{;j}{\cal G}_{\mu\nu}
\end{equation}

In contrast with \rf{embeddX},  the embedding  \rf{embeddY} does not
define a foliation as it refers to  a
a single surface, without free parameters.   Therefore  a  foliation still
needs to be constructed  and here this will be done by local
deformations of $\bar{\Sigma}$ as follows. 

A local perturbation with parameter $t$ of  a geometrical object $\bar{\Omega}$
in space-time is defined as  the   change of  $\bar{\Omega}$ under  the one
parameter group of 
diffeomorphisms generated by a vector field  $\zeta\;\;$:
$\Omega =\bar{\Omega}+\pounds_{t\zeta}\bar{\Omega}$ \cite{Geroch}.      
In particular the perturbation of the   vielbein $\bar{{\cal
Y}}^{\mu}_{,i}$  defined  by the solutions of 
\rf{embeddY} is
\begin{equation}
{\cal Z}^{\mu}_{,i}=\bar{{\cal Y}}^{\mu}_{,i}+\pounds_{t\zeta}
\bar{{\cal Y}}^{\mu}_{,i} =
\bar{{\cal Y}}^{\mu}_{,i}+t[\zeta, \bar{{\cal Y}}^{\alpha}_{,i}] .
\label{eq:h} 
\end{equation}
Thus, a  perturbation of  any tensor field defined in $\bar{\Sigma}$ may be  obtained by
taking its contraction with the  perturbed vielbein ${\cal Z}^{\mu}_{,i}$. 
A deformation of a  submanifold  $\bar{\Sigma}$  along  $\zeta$ is  a
perturbation of its 
geometry in that direction. A pure deformation corresponds
to the case of $\zeta$ orthogonal to  $\bar{\Sigma}$  \cite{Hojman}.

The  main difficulty associated with manifold deformations 
is the possible  existence of coordinate gauges. Indeed, 
different choices of $\zeta$  produce different
deformations\cite{Walker}. For example, taking a second transverse
vector  $\zeta'=(N,N'_{i})$, the deformation 
${\cal Z'}^{\mu}_{,i}=\bar{{\cal Y}}^{\mu}_{,i} +\pounds_{t{\zeta
'}}\bar{{\cal Y}}^{\mu}_{,i}$, differs from  \rf{h}  by
\begin{equation}  
{\cal Z'}^{\mu}_{,i}-{\cal Z}^{\mu}_{,i}= \pounds_{t(\zeta '
-\zeta)}\bar{{\cal Y}}^{\mu}_{,i} =t[\zeta' -\zeta,\bar{\cal Y}^{\mu}_{,i}]
\label{eq:deltah}  
\end{equation}
where $ \zeta ' -\zeta=(0, N'_{i}-N_{i})$. 
The two deformations become equal when the components $N_{i}$ and $N'_{i}$ 
are such that  the right hand side vanish.
In particular, this condition may result from a  mere
coordinate transformation in  $\bar{\Sigma}$. This means that a  deformation
may be generated (or destroyed) by a simple
coordinate transformation in $\bar{\Sigma}$. 
One example of this is given by the  ADM  foliation
\rf{embeddX} where a coordinate  transformation 
$x'^{i}=x^{i} +N^{i}$   generated by the shift
vector  $N^{i}$   may change or even  cancel  the foliation and hence
any notion of time associated with it. 
This  type of deformation does not belong
to the set of  physically admissible 
coordinate gauge independent (cgi for short) ones \cite{Bardeen}. 
However,  pure deformations  are  always
free of  coordinate gauges   because the 
propagation vector has  no components along $\bar{\Sigma}$ and this makes them
attractive to  the definition of time.  
In what follows  we will use pure deformations,   taking  $\zeta =\bar{{\cal
N}}$ in \rf{h},  to generate  a foliation  parametrized by $t$,
given by 
\begin{equation}
{\cal Z}^{\mu}_{,i}(x,t)=\bar{{\cal Y}}^{\mu}_{,i}(x)
+\pounds_{t{\cal N}}\bar{{\cal Y}}^{\mu}_{,i} =\bar{{\cal Y}}^{\mu}_{,i}(x)
+t\bar{{\cal N}}^{\mu}_{,i}(x) 
\label{eq:Zi}.
\end{equation}  
In this case, the  normal vector to the deformed hypersurface $\Sigma_{t}$ is
\[
{\cal N}^{\mu} =\bar{\cal N}^{\mu} =\pounds_{t\bar{\cal N}}\bar{\cal
N}^{\mu}=\bar{\cal N}^{\mu} 
+ t[\bar{\cal N}^{\mu},\bar{\cal N}^{\mu}] =\bar{\cal N}^{\mu}.
\]
Therefore each   leaf   $\Sigma_{t}$ of 
the foliation is described by the embedding coordinates (Notice that $t=0$, 
corresponds to the background  $\bar{\Sigma}$): 
\begin{equation}
{\cal Z}^{\mu} =\bar{{\cal Y}}^{\mu} +t\bar{{\cal N}}^{\mu}  \label{eq:Z}
\end{equation}
which must satisfy the  usual  embedding equations for each  leaf $\Sigma_{t}$:
\begin{equation}
h_{ij}={\cal Z}^{\mu}_{,i}{\cal Z}^{\nu}_{,j}{\cal G}_{\mu\nu},\;\;  
{\cal Z}^{\mu}_{,i}{\cal N}^{\nu}{\cal G}_{\mu\nu}=0,\;\;
{\cal N}^{\mu}{\cal N}^{\nu}{\cal G}_{\mu\nu}=-1. \label{eq:embeddZ}
\end{equation}
The   3-metric $h_{ij}$  may be
calculated  exactly from \rf{embeddZ}  in terms of the extrinsic curvature of
$\bar{\Sigma}$ :  
\begin{equation}
h_{ij}={\cal Z}^{\mu}_{,i}{\cal Z}^{\nu}_{,j}{\cal G}_{\mu\nu}=\bar{h}_{ij}
-2t\bar{b}_{ij} +t^{2}\bar{h}^{mn}\bar{b}_{im}\bar{b}_{jn}. \label{eq:gij}
\end{equation}
On the other hand, using the matrix notation $\bar{\bf h}=(\bar{h}_{mn})$  and
${\bf h}=(h_{mn})$ 
for the covariant metrics of $\bar{\Sigma}$ and $\Sigma_{t}$ respectively and
${\bf b}=(b_{mn})$, the  inverse of this metric may  be expressed
to any  order of  approximation $k$ as
\[
\stackrel{(k)}{{\bf h}^{-1}}=\left( \sum_{n=0}^{k} (\bar{\bf g}^{-1}{\bf
b})^{n} 
\right)^{2}\bar{\bf h}^{-1},\;\;\; \; {\bf h} \stackrel{(k)}{{\bf h}^{-1}}\approx 1 +0(t^{k+1}) .
\]
To complete our foliation we need to guarantee that  the deformations
described by  \rf{Z} in fact exist as isometrically  embedded  3-surfaces. This
is given by the  
fundamental theorem of hypersurfaces, stating that
for  a given pair of
tensors  $h_{ij}$ and  $b_{ij}$ satisfying the  conditions,
\begin{eqnarray}
R_{ijkl}&=&-2b_{i[k}b_{l]j} + {\cal R}_{\alpha\beta\gamma\delta} 
{\cal Z}^{\alpha}_{,i} {\cal Z}^{\beta}_{,j} {\cal Z}^{\gamma}_{k} {\cal Z}^{\delta}_{,l}
\label{eq:G1},\vspace{4mm}\\
2\nabla_{[k}b_{j]i} &=& {\cal R}_{\alpha\beta\gamma\delta}
{\cal Z}^{\alpha}_{,i} {\cal Z}^{\gamma}_{,j} {\cal Z}^{\delta}_{,k} {\cal N}^{\beta}
\label{eq:C1},
\end{eqnarray}  
then  there exists a hypersurface  $\Sigma_{t}$ embedded  in  ${\cal V}_{4}$
described by  ${\cal Z}^{\mu}$ and with normal vector  ${\cal N}^{\mu}$.

Therefore, by solving  \rf{G1} and \rf{C1}  for  a given
space-time metric ${\cal G}_{\alpha\beta}$  and  for  a  given pair of 
symmetric 3-tensors $h_{ij}$ and $b_{ij}$ we obtain each leaf of the  foliation
and its propagation vector ${\cal N}$.  Consequently,
\rf{G1} and \rf{C1} assume a fundamental role  in our formulation,
characterizing time  prior to any dynamical considerations.  It is  important
to notice they  are tensor equations,   so that  the foliation and the
consequent time is independent of  coordinates. 
The required tensors  $h_{ij}$ and  $b_{ij}$
are given by the dynamical equations  described in the following section. 

\section{Dynamics}
The use of  deformations  of embedded hypersurfaces to  generate a
time defining foliation,  where the embedding coordinates are taken
as the dynamical variables is not new (see for example \cite{Isham,Kuchar}. 
Here we take a  different approach,
where the  dynamical variables are still the  usual metric and momentum
tensors of the  hypersurface, but
contrarily to the traditional  procedures  we calculate the 
Dynamical equations  directly from the  equations  \rf{G1} and \rf{C1},  which
define the foliation. 
From the first two equations \rf{embeddZ}  we obtain ${\cal
G}^{\mu\nu}=h^{ij}{\cal Z}^{\mu}_{i}{\cal Z}^{\nu}_{j} +\Psi^{\mu\nu}$, where
$\Psi^{\mu\nu}$ is a symmetric tensor which is determined by  the remaining two
equations. It follows that $\Psi^{\mu\nu}=-{\cal N}^{\mu}{\cal N}^{\nu}$, so that 
\begin{equation}
{\cal G}^{\mu\nu}={\cal Z}^{\mu}_{,i}{\cal Z}^{\nu}_{,j}h^{ij}
-{\cal N}^{\mu}{\cal N}^{\nu}. \label{eq:Psi}
\end{equation}
Using this  expression in  Gauss' equations \rf{G1} we obtain
\begin{equation}
R_{jk}=h^{il}R_{ijkl}=-(b^{lk}b_{lj}-\mu b_{kj})+{\cal R}_{\beta\gamma}
{{\cal Z}}^{\beta}_{;j} {{\cal Z}}^{\gamma}_{;k}
+{\cal R}_{\alpha\beta\gamma\delta} 
{\cal N}^{\alpha} {\cal N}^{\delta}  {\cal Z}^{\beta}_{,j} {\cal
Z}^{\gamma}_{k}  
\label{eq:Rij} 
\end{equation}
and 
\begin{equation}
R=-(\kappa^{2} -\mu^{2})+{\cal R} +2{\cal R}_{\alpha\delta}{\cal
N}^{\alpha}{\cal N}^{\delta},  \label{eq:R}   
\end{equation}
where we have denoted  $\kappa =b^{ij}b_{ij}$ and  $\mu= h^{ij}b_{ij}$ is the
mean  curvature of  $\Sigma_{t}$.
Noting that  $\sqrt{-{\cal G}}=\sqrt{h}$, the  Einstein-Hilbert Lagrangian
for ${\cal V}_{4}$  is 
\begin{equation} 
{\cal L}={\cal R}\sqrt{-{\cal G}}=\left[ {R}- (\kappa^{2}-\mu^{2})+
 2{\cal R}_{\alpha\beta}{\cal N}^{\alpha}{\cal N}^{\beta}\right]
\sqrt{h}\label{eq:L} .
\end{equation}
This  expression is valid in any coordinate system  and in principle  we could
write the Hamiltonian and the  canonical equations. However, this would not be
practical. As in the  ADM case we   may  use  the  reference frame 
adapted to each leaf  $\Sigma_{t}$,  where  ${\cal N}^{\mu}
=\delta^{\mu}_{0}$.  In this frame the  second fundamental frame 
is  given by York's expression\cite{York}:
\begin{equation}
{b}_{ij}= \frac{1}{2}\dot{{h}}_{ij}. \label{eq:York}
\end{equation}
Then, by a direct calculation we find in this system
\[
{\cal R}_{\alpha\beta}{\cal N}^{\alpha}{\cal N}^{\beta} =
\Gamma^{\alpha}_{0\alpha,0}-\Gamma^{\alpha}_{00,\alpha}+
\Gamma^{\beta}_{0\alpha}\Gamma^{\alpha}_{\beta 0} 
-\Gamma^{\alpha}_{00}\Gamma^{\beta}_{\alpha\beta} 
= \kappa^{2}-\dot{\mu}.
\]
Therefore  the Lagrangian \rf{L}  is  equivalent to 
\[
{\cal L}={\cal R}\sqrt{-{\cal G}}=\left[ {R}- (\kappa^{2}+\mu^{2})+ 2\dot{\mu}
\right] \sqrt{h} .
\]
As usual, the   momentum canonically  conjugate to $h_{ij}$ is  defined
by the functional derivative 
\begin{equation}
p^{ij}=\frac{\delta {\cal L}}{\delta \dot{h}_{ij}}=  -(b^{ij} -\mu
h^{ij})\sqrt{h} \label{eq:p} 
\end{equation}
which has the same appearance as the ADM momentum  \rf{pi}. We also have the useful relations:
\[
p^{ij}p_{ij}=(\kappa^{2} +\mu^{2} )h,\;\;\; p=h_{ij}p^{ij}
=2\mu \sqrt{h}.
\]
Using \rf{p}, the second form may  be  expressed in terms of the momentum as
\[
b^{ij} = \frac{1}{\sqrt{h}}(p^{ij} -\frac{p}{2}g^{ij})
\]
and again using \rf{York},
\[
\dot{h}^{ij}=\frac{-2}{\sqrt{h}}(p^{ij}-\frac{p}{2}h^{ij}),\;\;\;
\dot{h}=
\frac{1}{2\sqrt{h}} (\dot{p}+\frac{p^{2}}{2\sqrt{h}}).
\]
Therefore, \rf{L}  may be expressed as
\begin{equation}
{\cal L}= R\sqrt{h}+\frac{1}{\sqrt{h}}(\frac{p^{2}}{2}-p^{ij}p_{ij}) +\dot{p},
\label{eq:L1}
\end{equation}
where we notice the absence  of  surface terms, meaning that we do not have to
worry about the compactness of $\Sigma_{t}$. On the other hand, the term in 
 $\dot{p}$  correspond to the same total time derivative of the ADM
 Lagrangian and  may be removed by canonical transformations
 \cite{Kuchar}, producing the effective Lagrangian:
\begin{equation}
{\cal L}_{eff}= R\sqrt{h}+\frac{1}{\sqrt{h}}(\frac{p^{2}}{2}-p^{ij}p_{ij}).
\end{equation}
The effective    Hamiltonian  follows directly from the Legendre transformation:
\begin{equation}
{\cal H}_{eff}=p^{ij}\dot{h}_{ij}-{\cal L}_{eff}=
-R\sqrt{h} +\frac{1}{\sqrt{h}}(\frac{p^{2}}{2}-p^{ij}p_{ij}).
\label{eq:newH}
\end{equation} 
As  we see,  the only dynamical variables present  are the metric
$h_{ij}$ and 
the momentum $p^{ij}$.

Hamilton's equations  may now be  calculated directly from \rf{newH}, without
much difficulty
\begin{eqnarray}
\dot{h}_{ij}&=&\frac{\delta {\cal H}_{eff}}{\delta p^{ij}}=
\frac{-2}{\sqrt{h}}\left(p_{ij}-\frac{p}{2}h_{ij} \right),
\vspace{3mm}\\
\dot{p}^{ij}&=&-\frac{\delta {\cal H}_{eff}}{\delta h_{ij}}= 
 (R^{ij}-\frac{1}{2}Rh_{ij})\sqrt{h} + \frac{1}{\sqrt{h}}\left[-p p^{ij} + 2p^{im}p^{j}_{m}
 -\frac{1}{2}(\frac{p^{2}}{2}-p^{mn}p_{mn})h^{ij}\right].
\end{eqnarray}
These are the same expressions  obtained by the 
time derivatives of  $h_{ij}$ and  $p^{ij}$ previously obtained by the
perturbation process. They also  coincide with the equations  derived from
the Euler-Lagrange equations  in the ADM formalism, when we take $N=1$ and  $
N_{i}=0$ so that they are  the correct equations.
Since they  were derived from the
Einstein-Hilbert  action \rf{L},  they  correspond  six of the  Einstein's
equations for the space-time\cite{ADM}. 
The remaining  four equations correspond to the two constraints
\rf{constraint1}, \rf{constraint2} of the ADM formalism (which  are equivalent
to  Einstein's equations 
$G_{00}=0 $ and   $G_{0i}=0$ for  the case of  pure 
gravitational field.). Here they are  not obtained from  the action because
the  variables $N$ and  $N^{i}$ were eliminated. Instead, the corresponding four
equations are derived from  the same integrability conditions
\rf{G1} and \rf{C1}. To keep up  with the analogy with the 
 ADM formulation, we may  define the new super Hamiltonian by (using 
\rf{p}) 
\begin{equation}
{\cal H}^{0}=\left[ R-(\kappa^{2}+\mu^{2})\right]\sqrt{h}=\left[ R
-\frac{p^{ij}p_{ij}}{h}\right] \sqrt{h},
\label{eq:H0} 
\end{equation}
Therefore, from \rf{R}  the  ADM constraint equation \rf{constraint1}  is
replaced by the equation 
\begin{equation}
{\cal H}_{0} ={\cal R}\sqrt{-{\cal G}}  
\end{equation}
On the other hand,   denoting   ${\cal H}_{i}=  \nabla^{k} b_{ki} -\mu_{,i}$,
then the contraction of   \rf{C1} with $h^{kl}$ gives the  equation
corresponding to  \rf{constraint2}
\begin{equation}
{\cal H}_{i}=- 2{\cal R}_{\mu\nu}{\cal Z}^{\mu}_{,i}{\cal N}^{\nu}
\label{eq:Hi} 
\end{equation}
These equations look  different from the corresponding  ADM
constraints because  we have a different deformation   and  no  vacuum
condition was imposed.  
Contrarily to the  ADM  case \rf{Hi} says that  the  momentum $p^{ij}$  is  not
paralelly transported  along $\Sigma_{t}$ and this is  a consequence of the
different slicing of the  space-time is foliated as compared with the ADM
foliation  \cite{Hojman}.
It is also worth noticing that   ${\cal H}_{i}$  does not depend on  $\dot{p}$,
meaning that the Codazzi constraint \rf{C1}  does not interfere   with the
removed total time derivative in the Hamiltonian.

Since  \rf{G1},\rf{C1}  were constructed  before any dynamical considerations,
they may be interpreted as  primary  constraints, in the sense that they hold
independently and  precede  the equations of  motion. This is  a  new and
more consistent situation as compared with the usual ADM formulation where the
constraints   appear  concomitantly with the  equations of motion. Therefore
we may implement different  situations on those equations  before  applying the
dynamical 
equations.  For example we may study the  pure gravitational case by
taking ${\cal R}_{\alpha\beta}=0 $ in \rf{H0} and  \rf{Hi}, obtaining the same
ADM constraints 
${\cal H}^{0}=0$,   ${\cal H}_{i}=0$ and yet ${\cal H}_{eff}\neq 0$.

\section{Discussion}
Our discussion  is limited by the  classical treatment given to the
problem. This means that there is no intention to  solve the
problem of time here, but simple  to clarify some of its  classical aspects.
In contrast to the cases of  Newtonian mechanics and  special relativity,
there is  no specific symmetry 
in general relativity  which  characterize a  time defining
foliation so that it  needs to be constructed by hand. 
In the ADM 3+1 splitting  the  foliation turns out to be inadequate for
the purposes of defining time because  it leads to  two time concepts
which are not necessarily compatible. 
As  an alternative, we have
introduced in space-time a hypersurface orthogonal  
foliation, with zero  shift and  fixed lapse. As usual, this was
initially constructed by 
pure deformations of a  given initial  3-surface. However, instead of  using
the foliation coordinates, as suggested by many authors, here we have
explored in full the integrability  conditions for the embedding of each
individual  leaf of the foliation. Since these are  tensor equations, the
resulting foliation and consequently the associated time is 
coordinate independent  of any specific symmetry of the
space-time.

With  the use of orthogonal foliation the concomitance problem of
the ADM formulation  was
eliminated and only the  foliation time  remains  defined prior to
any  dynamical considerations. 
The use of integrability conditions also  lead to the correct canonical
equations  from  a  non-zero  Hamiltonian.
We also notice the absence of  surface terms in the 
Hamiltonian, so that in general we do not require  compact
3-surfaces. The six equations \rf{newH} are complemented by the four
constraints \rf{H0} and \rf{Hi} which  may be  adapted to  different matter
distributions and in the vacuum they  reproduce the same ADM constraints.

A remark on the generality of this formulation should be added. 
Since  general relativity is characterized by the Einstein-Hilbert action, the
application of particular symmetries before the action principle would
certainly  create a particular situation. In this  formulation no specific
symmetry was applied to the Lagrangian, at least  up to  the expression
\rf{L}. Once we have a space-time characterized by Einstein's equations
derived from that Lagrangian,  then it is foliated  to  make  time and 
Hamilton's equations explicit. Therefore, in principle the above formulation
may apply to all space-times  which can be locally and orthogonally foliated. 

The  resulting foliation time  is  not necessarily  a  coordinate time.
Usually  a space-time metric is given  in a specific coordinate system which is
adapted to the symmetry  properties  of that space-time.
In the particular case of static or stationary  space-times and some
cosmological models, the foliation time is easily  identified with one of the
coordinates.  However such identification of  a coordinate with
time  is  only a mathematical convenience which should not
be confused with the definition of time itself.

\end{document}